\documentstyle[fleqn,run2col]{article}

\newcommand{\AmS}{{\protect\the\textfont2
  A\kern-.1667em\lower.5ex\hbox{M}\kern-.125emS}}
\newcommand{\msbar}{$\overline{\rm{MS}}$ }
\newcommand{\lsim}{\buildrel < \over {_\sim}}

\newcommand{\shat}{\hat{s}^2}
\newcommand{\chat}{\hat{c}^2}
\newcommand{\rwhat}{\Delta\hat{r}_W}
\newcommand{\rzhat}{\Delta\hat{r}_Z}
\newcommand{\dgama}{\hat\Delta_\gamma}

\newcommand{\etal}{{\em et al.}}
\newcommand{\be}{\begin{equation}}
\newcommand{\ee}{\end{equation}}
\newcommand{\ba}{\begin{array}}
\newcommand{\ea}{\end{array}}

\title{{\large\hfill UPR--885--T} \\
Global Fits to Electroweak Data Using GAPP}

\author{J. Erler\address{Department of Physics and Astronomy,
        University of Pennsylvania, \\ 
        David Rittenhouse Laboratory, 209 S. 33rd Street, 
        Philadelphia, PA 19104-6396}}

\begin{document}

\begin{abstract}
At Run II of the Tevatron it will be possible to measure the $W$ boson mass 
with a relative precision of about $2\times 10^{-4}$, which will eventually 
represent the best measured observable beyond the input parameters of the SM.  
Proper interpretation of such an ultrahigh precision measurement, either 
within the SM or beyond, requires the meticulous implementation and control 
of higher order radiative corrections.  The FORTRAN package GAPP, described 
here, is specifically designed to meet this need and to ensure the highest 
possible degrees of accuracy, reliability, adaptability, and efficiency.
\end{abstract}

\maketitle

\section{PRECISION TESTS}

Precision analysis of electroweak interactions follows three major objectives: 
high precision tests of the SM; the determination of its fundamental 
parameters; and studies of indications and constraints of possible new physics 
beyond the SM, such as supersymmetry or new gauge bosons. Currently, 
the experimental information comes from the very high precision $Z$ boson 
measurements at LEP~1 and the SLC, direct mass measurements and constraints 
from the Tevatron and LEP~2, and low energy precision experiments, such as in 
atomic parity violation, $\nu$ scattering, and rare decays.
These measurements are compared with the predictions of the SM and its 
extensions. The level of precision is generally very high. Besides the need 
for high-order loop calculations, it is important to utilize efficient 
renormalization schemes and scales to ensure sufficient convergence of 
the perturbative expansions. 

The tasks involved called for the creation of a special purpose FORTRAN 
package, GAPP, short for the {\it Global Analysis of Particle 
Properties}~\cite{GAPP}. It is mainly devoted to the calculation of 
pseudo-observables, i.e., observables appropriately idealized from 
the experimental reality. The reduction of raw data to pseudo-observables is 
performed by the experimenters with available packages (e.g., ZFITTER for $Z$ 
pole physics). For cross section and asymmetry measurements at LEP~2 (not 
implemented in the current version, GAPP$\_$99.7), however, this reduction is 
not optimal and convoluted expressions should be used instead. GAPP attempts to
gather all available theoretical and experimental information; it allows 
the addition of extra parameters describing new physics; it treats all relevant
SM inputs as global fit parameters; and it can easily be updated with new 
calculations, data, observables, or fit parameters. For clarity and speed it 
avoids numerical integrations throughout. It is based on the modified minimal 
subtraction ($\overline{\rm{MS}}$) scheme which demonstrably avoids large 
expansion coefficients. 

GAPP is endowed with the option to constrain nonstandard contributions to 
the {\em oblique parameters} defined to affect only the gauge boson 
self-energies~\cite{Peskin90} (e.g. $S$, $T$, and  $U$); specific anomalous 
$Z$ couplings; the number of active neutrinos (with standard couplings to the 
$Z$ boson); and the masses, mixings, and coupling strengths of extra $Z$ bosons
appearing in models of new physics. With view on the importance of 
supersymmetric extensions of the SM on one hand, and upcoming experiments on 
the other, I also included the $b \rightarrow s \gamma$ transition amplitude, 
and intend to add the muon anomalous magnetic moment. In the latter case, 
there are theoretical uncertainties from hadronic contributions which are 
partially correlated with the renormalization group (RG) evolutions of the QED 
coupling and the weak mixing angle. These correlations will be partially taken 
into account by including heavy quark effects in analytical form; see 
Ref.~\cite{Erler99} for a first step in this direction. By comparing this 
scheme with more conventional ones, it will also be possible to isolate a QCD 
sum rule and to rigorously determine the charm and bottom quark \msbar masses,
$\hat{m}_c$ and $\hat{m}_b$, with high precision. 

\section{GAPP}
\subsection{Basic structure}
In the default running mode of the current version, GAPP$\_$99.7, a fit is 
performed to 41 observables, out of which 26 are from $Z$ pole measurements at 
LEP~1 and the SLC. The Fermi constant, $G_F$ (from the muon lifetime), 
the electromagnetic fine structure constant, $\alpha$ (from the quantum Hall 
effect), and the light fermion masses are treated as fixed inputs. 
The exception is $\hat{m}_c$ which strongly affects the RG 
running\footnote{Quantities defined in the \msbar scheme are denoted by
a caret.} of $\hat\alpha (\mu)$ for $\mu > \hat{m}_c$. I therefore treat 
$\hat{m}_c$ as a fit parameter and include an external constraint with 
an enhanced error to absorb hadronic threshold uncertainties of other quark 
flavors, as well as theoretical uncertainties from the application of 
perturbative QCD at relatively low energies. Other fit parameters are 
the $Z$ boson mass, $M_Z$, the Higgs boson mass, $M_H$, the top quark mass,
$m_t$, and the strong coupling constant, $\alpha_s$, so that there are 
37 effective degrees of freedom. Given current precisions, $M_Z$ may 
alternatively be treated as an additional fixed input. 

The file {\tt fit.f} basically contains a simple call to the minimization
program MINUIT~\cite{MINUIT} (from the CERN program library) which is currently
used in data driven mode (see {\tt smfit.dat}). It in turn calls the core 
subroutine {\tt fcn} and the $\chi^2$-function {\tt chi2}, both contained in 
{\tt chi2.f}. Subroutine {\tt fcn} defines constants and flags; initializes 
parts of the one-loop package FF~\cite{Oldenborgh90,FF}; and makes the final 
call to subroutine {\tt values} in {\tt main.f} which drives the output 
(written to file {\tt smfit.out}). In {\tt chi2} the user actively changes and 
updates the data for the central values, errors, and correlation coefficients 
of the observables, and includes or excludes individual contributions to 
$\chi^2$ (right after the initialization, {\tt chi2 = 0.d0}). To each 
observable (as defined at the beginning of {\tt chi2}) corresponds an entry in 
each of the fields {\tt value, error, smval}, and {\tt pull}, containing 
the central observed value, the total (experimental and theoretical) error, 
the calculated fit value, and the standard deviation, respectively. 
The function {\tt chi2} also contains calls to various other subroutines where 
the actual observable calculations take place. These are detailed in 
the following subsections. 

Another entry to GAPP is provided through {\tt mh.f} which computes 
the probability distribution function of $M_H$. The probability distribution
function is the quantity of interest within Bayesian data analysis (as opposed 
to point estimates frequently used in the context of classical methods), and
defined as the product of a prior density and the likelihood, 
${\cal L} \sim \exp( - \chi^2/2)$. If one chooses to disregard any further
information on $M_H$ (such as from triviality considerations or direct 
searches) one needs a {\em non-informative prior}. It is recommanded to choose 
a flat prior in a variable defined on the whole real axis, which in the case of
$M_H$ is achieved by an equidistant scan over $\log M_H$. An {\em informative
prior\/} is obtained by activating one of the approximate Higgs exclusion 
curves from LEP~2 near the end of {\tt chi2.f}. These curves affect values of 
$M_H$ even larger than the corresponding quoted 95\% CL lower limit and 
includes an extrapolation to the kinematic limit; notice that this corresponds 
to a conservative treatment of the upper $M_H$ limit. 

Contour plots can be obtained using the routine {\tt mncontours} from MINUIT. 
For the cases this fails, some simpler and slower but more robust contour 
programs are also included in GAPP, but these have to be adapted by the user 
to the case at hand. 

\subsection{$\hat\alpha$, $\sin^2\hat\theta_W$, $M_W$}
At the core of present day electroweak analyses is the interdependence between
$G_F$, $M_Z$, the $W$ boson mass, $M_W$, and the weak mixing angle, 
$\sin^2\theta_W$. In the \msbar scheme it can be written 
as~\cite{Sirlin90,Degrassi91},
\be
  \shat = \frac{A^2}{M_W^2 (1 - \rwhat)}, \hspace{15pt}
  \shat\chat = \frac{A^2}{M_Z^2 (1 - \rzhat)},
\ee
where,
\be
  A=\left[ \frac{\pi\alpha}{\sqrt{2} G_F} \right]^{1/2}=37.2805(2)\hbox{ GeV},
\ee
$\shat$ is the \msbar mixing angle, $\chat = 1 -\shat$, and where,
\be
  \rwhat = {\alpha\over \pi} \dgama + 
  \frac{\hat\Pi_{WW}(M_W^2) - \hat\Pi_{WW}(0)}{M_W^2} + \mbox{V + B},
\ee
and,
\be
  \rzhat = \rwhat + (1 - \rwhat)
  \frac{\hat\Pi_{ZZ}(M_Z^2) - {\hat\Pi_{WW}(M_W^2)\over \chat}}{M_Z^2},
\ee
collect the radiative corrections computed in {\tt sin2th.f}. The $\hat\Pi$ 
indicate \msbar subtracted self-energies, and V + B denote the vertex and 
box contributions to $\mu$ decay. Although these relations involve the \msbar 
gauge couplings they employ on-shell gauge boson masses, absorbing a large 
class of radiative corrections~\cite{Degrassi99}.

$\rwhat$ and $\rzhat$ are both dominated by the contribution $\dgama(M_Z)$
which is familiar from the RG running of the electromagnetic coupling,
\be
  \hat\alpha (\mu) = {\alpha\over 1 - {\alpha\over\pi} \dgama (\mu)},
\ee
and computed in {\tt alfahat.f} up to four-loop ${\cal O}(\alpha\alpha_s^3)$.
Contributions from $c$ and $b$ quarks are calculated using an unsubtracted 
dispersion relation~\cite{Erler99}. If $\mu$ is equal to the mass of a quark,
three-loop matching is performed and the definition of $\hat\alpha$ changes
accordingly. Pure QED effects are included up to next-to-leading order (NLO)
while higher orders are negligible. Precise results can be obtained for 
$\mu < 2 m_\pi$ and $\mu \geq m_c$. 

Besides full one-loop electroweak corrections, $\rwhat$ and $\rzhat$ include
enhanced two-loop contributions of ${\cal O}(\alpha^2 m_t^4)$~\cite{Barbieri92}
(implemented using the analytic expressions of Ref.~\cite{Fleischer93}) and 
${\cal O} (\alpha^2 m_t^2)$~\cite{Degrassi96} (available as  expansions in
small and large $M_H$); mixed electroweak/QCD corrections of 
${\cal O} (\alpha\alpha_s)$~\cite{Djouadi87} and
${\cal O} (\alpha\alpha_s^2 m_t^2)$~\cite{Chetyrkin95}; the analogous
mixed electroweak/QED corrections of ${\cal O} (\alpha^2)$; and fermion mass 
corrections also including the leading gluonic and photonic corrections.

\subsection{$Z$ decay widths and asymmetries}
\label{sec:zdecays}
The partial width for $Z \rightarrow f\bar{f}$ decays is given by,
$$
  \Gamma_{f\bar{f}} = {N_C^f M_Z\hat\alpha\over 24\shat\chat} |\hat\rho_f|
    \left[1 - 4 |Q_f|   {\rm Re} (\hat\kappa_f) \shat 
            + 8  Q_f^2 \hat{s}^4 |\hat\kappa_f|^2 \right] 
$$
\be
  \times \left[1 + \delta_{\rm QED} + \delta_{\rm QCD}^{\rm NS} + 
  \delta_{\rm QCD}^{\rm S} - {\hat\alpha\hat\alpha_s\over 4\pi^2} Q_f^2 + 
  {\cal O} (m_f^2) \right].
\label{eq:gammaz}
\ee
$N_C^f$ is the color factor, $Q_f$ is the fermion charge, and $\hat\rho_f$ 
and $\hat\kappa_f$ are form factors which differ from unity through one-loop 
electroweak corrections~\cite{Degrassi91A} and are computed in {\tt rho.f} and
{\tt kappa.f}, respectively. For $f \neq b$ there are no corrections of 
${\cal O} (\alpha^2 m_t^4)$ and contributions of ${\cal O} (\alpha^2 m_t^2)$ 
to $\hat\kappa_f$~\cite{Degrassi97} and $\hat\rho_f$~\cite{Degrassi99} are 
very small and presently neglected. On the other hand, vertex corrections of 
${\cal O} (\alpha\alpha_s)$~\cite{Czarnecki97} are important and shift 
the extracted $\alpha_s$ by $\sim 0.0007$.

The $Z \rightarrow b\bar{b}$ vertex receives extra corrections due to heavy top
quark loops. They are large and have been implemented in {\tt bvertex.f} based 
on Ref.~\cite{Bernabeu91}. ${\cal O}(\alpha^2 m_t^4)$ 
corrections~\cite{Barbieri92,Fleischer93} are included, as well, while those of
${\cal O} (\alpha^2 m_t^2)$ are presently unknown. The leading QCD effects of 
${\cal O} (\alpha\alpha_s m_t^2)$~\cite{Fleischer92} and all subleading
${\cal O} (\alpha\alpha_s)$ corrections~\cite{Harlander98} are incorporated 
into $\hat\rho_b$ and $\hat\kappa_b$, but not the 
${\cal O} (\alpha\alpha_s^2 m_t^2)$ contribution which is presently available 
only for nonsinglet diagrams~\cite{Chetyrkin99}. 

In Eq.~(\ref{eq:gammaz}), $\delta_{\rm QED}$ are the ${\cal O} (\alpha)$ and
${\cal O} (\alpha^2)$ QED corrections. $\delta_{\rm QCD}^{\rm NS}$ are 
the universal QCD corrections up to ${\cal O} (\alpha_s^3)$ which include 
quark mass dependent contributions due to double-bubble type 
diagrams~\cite{Chetyrkin96,Larin95}. $\delta_{\rm QCD}^{\rm S}$ are 
the singlet contributions to the axial-vector and vector partial widths 
which start, respectively, at ${\cal O}(\alpha_s^2)$ and 
${\cal O} (\alpha_s^3)$, and induce relatively large family universal but 
flavor non-universal $m_t$ effects~\cite{Larin95,Kniehl90}. The corrections
appearing in the second line of Eq.~(\ref{eq:gammaz}) are evaluated in
{\tt lep100.f}.

The dominant massless contribution to $\delta_{\rm QCD}^{\rm NS}$ can be 
obtained by analytical continuation of the Adler $D$-function, which 
(in the \msbar scheme) has a very well behaved perturbative expansion 
$\sim 1 + \sum_{i=0} d_i a_s^{i+1}$ in $a_s = \hat\alpha_s (M_Z)/\pi$ (see 
the Appendix for details). The process of analytical continuation from 
the Euclidean to the physical region induces further terms which are 
proportional to $\beta$-function coefficients, enhanced by powers of 
$\pi^2$, and start at ${\cal O} (\alpha_s^3)$. Fortunately, these 
terms~\cite{Kataev95} involve only known coefficients up to 
${\cal O} (\alpha_s^5)$, and the only unknown coefficient in 
${\cal O} (\alpha_s^6)$ is proportional to the four-loop Adler function 
coefficient, $d_3$. In the massless approximation,
\be
 \delta_{\rm QCD}^{\rm NS}\approx a_s + 1.4092 a_s^2 - (0.681 + 12.086) a_s^3 +
\label{eq:qcdns}
\ee
$$
  (d_3 - 89.19) a_s^4 + (d_4 + 79.7) a_s^5 + (d_5 - 121 d_3 + 3316) a_s^6,
$$
and terms of order $a_s^7 \sim 10^{-10}$ are clearly negligible. Notice, that
the ${\cal O} (\alpha_s^6)$ term effectively {\em reduces\/} the sensitivity 
to $d_3$ by about 18\%. Eq.~(\ref{eq:qcdns}) amounts to a reorganization of 
the perturbative series in terms of the $d_i$ times some function of 
$\alpha_s$; a similar idea is routinely applied to the perturbative QCD 
contribution to $\tau$ decays~\cite{Diberder92}. 

Final state fermion mass effects~\cite{Chetyrkin96,Chetyrkin97} of 
${\cal O}(m_f^2)$ (and ${\cal O} (m_b^4)$ for $b$ quarks) are best evaluated by
expanding in $\hat{m}^2_q (M_Z)$ thus avoiding large logarithms in the quark 
masses. The singlet contribution of ${\cal O} (\alpha_s^2 m_b^2)$ is also 
included. 

The dominant theoretical uncertainty in the $Z$ lineshape determination of 
$\alpha_s$ originates from the massless quark contribution, and amounts to 
about $\pm 0.0004$ as estimated in the Appendix. There are several further 
uncertainties, all of ${\cal O} (10^{-4})$: from the ${\cal O} (\alpha_s^4)$  
heavy top quark contribution to the axial-vector part of 
$\delta_{\rm QCD}^{\rm S}$; from the missing ${\cal O}(\alpha\alpha_s^2 m_t^2)$
and ${\cal O} (\alpha^2 m_t^2)$ contributions to the $Zb\bar{b}$-vertex; 
from further non-enhanced but cohering ${\cal O}(\alpha\alpha_s^2)$-vertex 
corrections; and from possible contributions of non-perturbative origin.
The total theory uncertainty is therefore,
\be
\label{eq:alphastotal}
  \Delta\alpha_s (M_Z)= \pm 0.0005,
\ee
which can be neglected compared to the current experimental error. 
If $\hat{m}_b$ is kept fixed in a fit, then its parametric error would 
add an uncertainty of $\pm 0.0002$, but this would not change the total
uncertainty~(\ref{eq:alphastotal}).

Polarization asymmetries are (in some cases up to a trivial factor 3/4 or 
a sign) given by the asymmetry parameters,
\be
   A_f = \frac{1 - 4 |Q_f|   {\rm Re} (\hat\kappa_f) \shat}
              {1 - 4 |Q_f|   {\rm Re} (\hat\kappa_f) \shat 
                 + 8  Q_f^2 \hat{s}^4 |\hat\kappa_f|^2},
\ee
and the forward-backward asymmetries by,
\be
   A_{FB} (f) = {3\over 4} A_e A_f.
\ee
The hadronic charge asymmetry, $Q_{FB}$, is the linear combination,
\be
   Q_{FB} = \sum\limits_{q = d,s,b} R_q A_{FB}(q) - 
            \sum\limits_{q = u,c}   R_q A_{FB}(q),
\ee
and the hadronic peak cross section, $\sigma_{\rm had}$, is stored in 
{\tt sigmah}, and defined by,
\be
  \sigma_{\rm had}={12\pi\Gamma_{e^+e^-}\Gamma_{\rm had}\over M_Z^2\Gamma_Z^2}.
\ee   
Widths and asymmetries are stored in the fields {\tt gamma(f)}, {\tt alr(f)}, 
and {\tt afb(f)}. The fermion index, {\tt f}, and the partial width ratios, 
{\tt R(f)}, are defined in 
Table~\ref{tab:lep100}.
\begin{table}[h]
\caption{Some of the variables used in {\tt lep100.f}. $\Gamma_{\rm inv}$ and
$\Gamma_{\rm had}$ are the invisible and hadronic decays widths, respectively.}
\label{tab:lep100}
\begin{tabular}{|c|c|l|l|}
\hline
 0 & $\nu$   & {\tt gamma(0)} = $\Gamma_{\rm inv}$ & {\tt alr(0)} = 1 \\
 1 & $e$     & {\tt R(1)} = $\Gamma_{\rm had}/\Gamma_{e^+e^-}$ & 
               {\tt alr(1)} = $A_e$ \\
 2 & $\mu$   & {\tt R(2)} = $\Gamma_{\rm had}/\Gamma_{\mu^+\mu^-}$ & 
               {\tt alr(2)} = $A_\mu$ \\
 3 & $\tau$  & {\tt R(3)} = $\Gamma_{\rm had}/\Gamma_{\tau^+\tau^-}$ &
               {\tt alr(3)} = $A_\tau$ \\
 4 & $u$     & {\tt R(4)} = $\Gamma_{u\bar{u}}/\Gamma_{\rm had}$ & --- \\
 5 & $c$     & {\tt R(5)} = $\Gamma_{c\bar{c}}/\Gamma_{\rm had}$ & 
               {\tt alr(5)} = $A_c$ \\
 6 & $t$     & {\tt R(6)} = 0                                    & --- \\
 7 & $d$     & {\tt R(7)} = $\Gamma_{d\bar{d}}/\Gamma_{\rm had}$ & --- \\
 8 & $s$     & {\tt R(8)} = $\Gamma_{s\bar{s}}/\Gamma_{\rm had}$ & 
               {\tt alr(8)} = $A_s$ \\
 9 & $b$     & {\tt R(9)} = $\Gamma_{b\bar{b}}/\Gamma_{\rm had}$ & 
               {\tt alr(9)} = $A_b$ \\
10 & had     & {\tt gamma(10)} = $\Gamma_{\rm had}$ & {\tt afb(10)} $=Q_{FB}$\\
11 & all     & {\tt gamma(11)} = $\Gamma_Z$                      & --- \\
\hline
\end{tabular}
\end{table}

\subsection{Fermion masses}
I use \msbar masses as far as QCD is concerned, but retain on-shell masses 
for QED since renormalon effects are unimportant in this case. This results in 
a hybrid definition for quarks. Accordingly, the RG running of the masses 
to scales $\mu \neq \hat{m}_q$ uses pure QCD anomalous dimensions. The running 
masses correspond to the functions {\tt msrun($\mu$), mcrun($\mu$)}, etc. which
are calculated in {\tt masses.f} to three-loop order. Anomalous dimensions are 
also available at four-loop order~\cite{Vermaseren97}, but can safely be 
neglected. Also needed is the RG evolution of $\alpha_s$ which is implemented 
to four-loop precision~\cite{Ritbergen97} in {\tt alfas.f}.

I avoid pole masses for the five light quarks throughout. Due to renormalon 
effects, these can be determined only up to ${\cal O}(\Lambda_{\rm QCD})$ 
and would therefore induce an irreducible uncertainty of about 0.5~GeV. 
In fact, perturbative expansions involving the pole mass show unsatisfactory 
convergence. In contrast, the \msbar mass is a short distance mass which can, 
in principle, be determined to arbitrary precision, and perturbative
expansions are well behaved with coefficients of order unity (times group 
theoretical factors which grow only geometrically). Note, however, that 
the coefficients of expansions involving large powers of the mass, $\hat{m}^n$,
are rather expected to be of ${\cal O}(n)$. This applies, e.g., to decays of 
heavy quarks ($n=5$) and to higher orders in light quark mass expansions.

The top quark pole mass enters the analysis when the results on $m_t$ from 
on-shell produced top quarks at the Tevatron are included. In subroutine 
{\tt polemasses(nf,mpole)} $\hat{m}_q(\hat{m}_q)$ is converted to the quark
pole mass, {\tt mpole}, using the two-loop perturbative relation from 
Ref.~\cite{Gray90}. The exact three-loop result~\cite{Melnikov99} has been
approximated (for $m_t$) by employing the BLM~\cite{Brodsky83} scale for 
the conversion. Since the pole mass is involved it is not surprising that 
the coefficients are growing rapidly. The third order contribution is 31\%, 
75\%, and 145\% of the second order for $m_t$ (${\tt nf} = 6$), $m_b$ 
(${\tt nf} = 5$), and $m_c$ (${\tt nf} = 4$), respectively. I take 
the three-loop contribution to the top quark pole mass of about 0.5~GeV 
as the theoretical uncertainty, but this is currently negligible relative to 
the experimental error. At a high energy lepton collider it will be possible 
to extract the \msbar top quark mass directly and to abandon quark pole masses 
altogether. 

\subsection{$\nu$ scattering}

The ratios of neutral-to-charged current cross sections, 
\be
  R_\nu = \frac{\sigma_{\nu N}^{NC}}{\sigma_{\nu N}^{CC}}, \hspace{25pt}
  R_{\bar\nu} = \frac{\sigma_{\bar\nu N}^{NC}}{\sigma_{\bar\nu N}^{CC}},
\ee
have been measured precisely in deep inelastic $\nu$ ($\bar\nu$) hadron 
scattering (DIS) at CERN (CDHS and CHARM) and Fermilab (CCFR). The most precise
result was obtained by the NuTeV Collaboration at Fermilab who determined 
the Paschos-Wolfenstein ratio,
\be
  R^- = \frac{\sigma_{\nu N}^{NC} - \sigma_{\bar\nu N}^{NC}}
             {\sigma_{\nu N}^{CC} - \sigma_{\bar\nu N}^{CC}}
      \sim R_\nu - r R_{\bar\nu},
\ee
with $r = \sigma_{\bar\nu N}^{CC}/\sigma_{\nu N}^{CC}$. Results on $R_\nu$ are 
frequently quoted in terms of the on-shell weak mixing angle (or $M_W$) as this
incidentally gives a fair description of the dependences on $m_t$ and $M_H$.
One can write approximately,
\be
  R_\nu       = g_L^2 + g_R^2 r,        \hspace{6pt}
  R_{\bar\nu} = g_L^2 + {g_R^2\over r}, \hspace{6pt}
  R^-         = g_L^2 - g_R^2,
\ee
where,
\be
  g_L^2 = {1\over 2} - \sin^2\theta_W + {5\over9} \sin^4\theta_W, \hspace{10pt}
  g_R^2 =                               {5\over9} \sin^4\theta_W.
\ee
However, the study of new physics requires the implementation of the actual 
linear combinations of effective four-Fermi operator coefficients, 
$\epsilon_{L,R}(u)$ and $\epsilon_{L,R}(d)$, which have been measured. 
With the appropriate value for the average momentum transfer, {\tt q2}, 
as input, these are computed in the subroutines 
{\tt nuh(q2,epsu\_L,epsd\_L,epsu\_R,epsd\_R)} (according to 
Ref.~\cite{Marciano80}), {\tt nuhnutev}, {\tt nuhccfr}, and {\tt nuhcdhs}, all 
contained in file {\tt dis.f}. Note, that the CHARM results have been adjusted 
to CDHS conditions~\cite{Perrier95}. While the experimental correlations 
between the various DIS experiments are believed to be negligible, large 
correlations are introduced by the physics model through charm mass threshold 
effects, quark sea effects, radiative corrections, etc. 
I constructed the matrix of correlation coefficients using the analysis 
in Ref.~\cite{Perrier95},
$$
R^-\hspace{13pt}R_\nu\hspace{15pt}R_\nu\hspace{15pt}R_\nu\hspace{15pt} 
R_{\bar\nu}\hspace{15pt}R_{\bar\nu}\hspace{15pt}R_{\bar\nu}\hspace{23pt}
$$
\be
   \left( \ba{ccccccc} 
     1.00 & 0.10 & 0.10 & 0.10 & 0.00 & 0.00 & 0.00 \\
     0.10 & 1.00 & 0.40 & 0.40 & 0.10 & 0.10 & 0.10 \\
     0.10 & 0.40 & 1.00 & 0.40 & 0.10 & 0.10 & 0.10 \\
     0.10 & 0.40 & 0.40 & 1.00 & 0.10 & 0.10 & 0.10 \\
     0.00 & 0.10 & 0.10 & 0.10 & 1.00 & 0.15 & 0.15 \\
     0.00 & 0.10 & 0.10 & 0.10 & 0.15 & 1.00 & 0.15 \\
     0.00 & 0.10 & 0.10 & 0.10 & 0.15 & 0.15 & 1.00 \ea \right).
\ee

The effective vector and axial-vector couplings, $g_V^{\nu e}$ and 
$g_A^{\nu e}$, from elastic $\nu e$ scattering are calculated in subroutine 
{\tt nue(q2,gvnue,ganue)} in file {\tt nue.f}. The momentum transfer, 
${\tt q2}$, is currently set to zero~\cite{Marciano95}. Needed is the low 
energy $\rho$ parameter, {\tt rhonc}, which describes radiative corrections to 
the neutral-to-charged current interaction strengths. Together with 
{\tt sin2t0} (described below) it is computed in file {\tt lowenergy.f}.

\begin{table}[t]
\caption{Coefficients (\msbar) appearing in the $\beta$-function of a simple 
group~\cite{Ritbergen97}; in non-Abelian corrections to the QED 
$\beta$-function (denoted $\tilde{D}$)~\cite{Erler99,Chetyrkin97A}; 
in the Adler $D$-function~\cite{Surguladze91} (rescaled by an overall factor 
1/3); and in $R_{\rm had}$ (analytical continuation of $D$). The first four 
segments correspond, respectively, to the first four loop orders of non-singlet
type. The fifth segment is the singlet (double triangle) contribution in 
${\cal O}(\alpha_s^3)$. In $\tilde{D}$, $D$, and $R_{\rm had}$, an overall 
factor $\hat\alpha M_Z$ and the sums involving charges or $Z$ couplings have 
been dropped. The completely symmetrical tensors of rank four, $d_A$ and $d_F$,
as well as $T_F$ when appearing in parenthesis, apply to the $\beta$-function 
only. Each $T_F$ is understood to be multiplied by the number of flavors, 
$n_f$, except for the singlet term involving the symmetrical structure 
constants, $d$.}
\label{tab:coeff}
\begin{tabular}{|l|rrrr|}
\hline
group factor & $\beta$ & $\tilde{D}$ & $D$ & $R_{\rm had}$ \\
\hline\hline
$C_A$                            &  0.92 &  ---  &  ---  &  ---   \\
$(T_F)$                          &$-0.33$&$-0.33$&  0.33 &  0.33  \\
\hline
$C_A^2$                          &  0.71 &  ---  &  ---  &  ---   \\
$C_A T_F$                        &$-0.42$&  ---  &  ---  &  ---   \\
$C_F (T_F)$                      &$-0.25$&$-0.25$&  0.25 &  0.25  \\
\hline 
$C_A^3$                          &  0.83 &  ---  &  ---  &  ---   \\
$C_A^2 T_F$                      &$-0.82$&  ---  &  ---  &  ---   \\
$C_A C_F (T_F)$                  &$-0.36$&$-0.23$&  0.18 &  0.18  \\
$C_A T_F^2$                      &  0.09 &  ---  &  ---  &  ---   \\
$C_F^2 (T_F)$                    &  0.03 &  0.03 &$-0.03$&$-0.03$ \\
$C_F T_F (T_F)$                  &  0.08 &  0.08 &$-0.06$&$-0.06$ \\
\hline
$C_A^4$                          &  1.19 &  ---  &  ---  &  ---   \\
$C_A^3 T_F$                      &$-1.67$&  ---  &  ---  &  ---   \\
$C_A^2 C_F (T_F)$                &$-0.23$&$-0.28$&  0.32 &$-0.38$ \\
$C_A^2 T_F^2$                    &  0.50 &  ---  &  ---  &  ---   \\
$C_A C_F^2 (T_F)$                &$-0.42$&$-0.07$&  0.51 &  0.51  \\
$C_A C_F T_F (T_F)$              &  0.51 &  0.42 &$-0.71$&$-0.21$ \\
$C_A T_F^3$                      &  0.01 &  ---  &  ---  &  ---   \\
$C_F^3 (T_F)$                    &  0.18 &  0.18 &$-0.18$&$-0.18$ \\
$C_F^2 T_F (T_F)$                &$-0.17$&$-0.17$&  0.02 &  0.02  \\
$C_F T_F^2 (T_F)$                &  0.02 &  0.02 &  0.09 &$-0.01$ \\
\hline
$d_A^2/N_A$                      &  1.07 &  ---  &  ---  &  ---   \\
$d_A d_F/N_A$                    &$-2.38$&  ---  &  ---  &  ---   \\
$d_F^2/N_A$ $(T_F^2 d^2/4)$      &  0.50 &  0.50 &$-0.50$&$-0.50$ \\ 
\hline\hline
$\bar{y}$                        &$-0.02$&$-0.01$&  0.02 &$-0.01$ \\ 
$\Delta\mu$                      &  0.17 &  0.09 &  0.11 &  0.09  \\
\hline
$\sigma_0$                       &  0.83 &  0.28 &  0.37 &  0.31  \\
$\Delta\sigma$                   &  0.13 &  0.07 &  0.09 &  0.08  \\
\hline
\end{tabular}
\end{table}

\subsection{Low energy observables}
The weak atomic charge, {\tt Qw}, from atomic parity violation and fixed target
$ep$ scattering is computed in subroutine {\tt apv(Qw,Z,AA,C1u,C1d,C2u,C2d)}
where {\tt Z} and {\tt AA} are, respectively, the atomic number and weight.
Also returned are the coefficients from lepton-quark effective four-Fermi 
interactions which are calculated according to~\cite{Marciano84}. 

These observables are sensitive to the low energy mixing angle, {\tt sin2t0}, 
which defines the electroweak counterpart to the fine structure constant and is
similar to the one introduced in Ref.~\cite{Sirlin90}. There is significant 
correlation between the hadronic uncertainties from the RG evolutions of 
$\hat\alpha$ and the weak mixing angle. Presently, this correlation is ignored,
but with the recent progress in atomic parity violation experiments it should 
be accounted for in the future.

An additional source of hadronic uncertainty is introduced by $\gamma Z$-box 
diagrams which are unsuppressed at low energies. At present, this uncertainty
can be neglected relative to the experimental precision. 

Besides {\tt apv}, file {\tt pnc.f} contains in addition the subroutine 
{\tt moller} for the anticipated polarized fixed target M\o ller scattering 
experiment at SLAC. Radiative corrections are included following 
Ref.~\cite{Czarnecki96}.

\subsection{$b \rightarrow s \gamma$}

Subroutine {\tt bsgamma} returns the decay ratio,
\be
 R = \frac{{\cal B} (b \rightarrow s \gamma)}{{\cal B} (b \rightarrow c e\nu)}.
\ee
It is given by~\cite{Greub96,Borzumati98},
\be
  R = {6\alpha\over \pi} \left| {V_{ts}^* V_{tb}\over V_{cb}} \right|^2 
      {S\over f(z)} \frac{|\bar{D}|^2 + A/S + \delta_{NP} + \delta_{EW}}
                         {(1 + \delta_{NP}^{SL}) (1 + \delta_{EW}^{SL})},
\ee
where $|V_{ts}^* V_{tb}/V_{cb}|^2 = 0.950$ is a combination of 
Cabbibo-Kobayashi-Maskawa matrix elements and $S$ is the Sudakov 
factor~\cite{Kagan99}. $\delta_{NP}$ and $\delta_{EW}$ are non-perturbative and
NLO electroweak~\cite{Czarnecki98} corrections, both for 
the $b\rightarrow s\gamma$ and the semileptonic ($b \rightarrow c e\nu$) decay 
rates.
\be
  \bar{D}= C_7^0 + {\hat\alpha_s (\hat{m}_b) \over 4\pi} (C_7^1 + V),
\ee
is called the reduced amplitude for the process $b \rightarrow s \gamma$, and
is given in terms of the Wilson coefficient $C_7$ at NLO. $C_7$ and the other 
$C_i$ appearing below are effective Wilson coefficients with NLO RG 
evolution~\cite{Chetyrkin98} from the weak scale to $\mu = \hat{m}_b$ 
understood. The NLO matching conditions at the weak scale have been calculated 
in Ref.~\cite{Adel94}. $\bar{D}$ includes the virtual gluon corrections,
\be
\label{eq:vrtuel}
  V = r_2 C_2^0  + r_7 C_7^0 + r_8 C_8^0,
\ee
so that it squares to a positive definite branching fraction. On the other
hand, the amplitude for gluon Bremsstrahlung ($b \rightarrow s \gamma g$), 
$$
\ba{l}
  A = {\hat\alpha_s (\hat{m}_b) \over \pi} 
  [C_2^0 (C_8^0 f_{28}(1) + C_7^0 f_{27}(1) + C_2^0 f_{22}(1)) + 
\ea
$$
\be
\label{eq:abrems}
\hspace{10pt} C_8^0 (C_8^0 f_{88}(\delta) + C_7^0 f_{78}(\delta))
                             + (C_7^0)^2 f_{77}(\delta)],
\ee
is added linearly to the cross section. The Wilson coefficient $C_2^0$ is 
defined as in Ref.~\cite{Buras94}. It enters only at NLO, is significantly 
larger than $C_7^0$, and dominates the NLO contributions. The parameter 
$0 \leq \delta \leq 1$ in the coefficient functions $f_{ij}$ characterizes 
the minimum photon energy and has been set to $\delta = 0.9$~\cite{Kagan99}, 
except for the first line in Eq.~(\ref{eq:abrems}) where $\delta = 1.0$ 
corresponding to the full cross section. The $f_{2i}$ are complicated
integrals which can be solved in terms of polylogarithms up to 5th order.
In the code I use an expansion in $z = m_c^2/m_b^2$ and $\delta = 1.0$.
Once experiments become more precise the correction to $\delta = 0.9$ should
be included. 

$f(z)$ is the phase space factor for the semileptonic decay rate including
NLO corrections~\cite{Nir89}. I defined the \msbar mass ratio in  
$z = [\hat{m}_c (\hat{m}_b)/\hat{m}_b(\hat{m}_b)]^2$ at the common scale, 
$\mu = \hat{m}_b$, which I also assumed for the factor $\hat{m}_b^5$ 
multiplying the decay widths. Since I do not reexpand the denominator this 
effects the phase space function at higher orders. Using 
the ${\cal O}(\alpha_s^2)$-estimate\footnote{I computed the 
${\cal O}(\alpha_s^2)$ coefficient for comparison only, and did not include it 
in the code.} from Ref.~\cite{Czarnecki98A}, I obtain for the semileptonic 
decay width,
\be
\label{eq:slas2}
  \Gamma_{SL} \sim
  \hat{m}_b^5 f_0(z) \left[1 + 2.7 {\hat\alpha_s (\hat{m}_b) \over \pi} - 
                          1.6 \left({\hat\alpha_s \over \pi}\right)^2\right],
\ee
where $f_0(z)$ is the leading order phase space factor. It is amusing that 
the coefficients in Eq.~(\ref{eq:slas2}) are comfortably (and perhaps somewhat 
fortuitously) small, with the ${\cal O}(a_s^2)$-coefficient even smaller than 
the one in Ref.~\cite{Czarnecki98A} where a low scale running mass had been 
advocated. Moreover, using the prefactor $\hat{m}_b^5$ in the numerator of $R$
reduces the size of $r_7$ in Eq.~(\ref{eq:vrtuel}) and therefore 
the coefficient $\kappa (\delta) = f_{77} (\delta) + r_7/2$ which multiplies 
the term $a_s (C_7^{0,{\rm eff}})^2$. I obtain $-2.1 < \kappa (\delta) < 1.4$, 
while with the pole mass prefactor $M_b^5$ one would have 
$-8.7 \leq \kappa (\delta) < -5.3$.

\section*{Acknowledgements}
It is a pleasure to thank Paul Langacker and Damien Pierce for collaborations
on precision analyses. I am grateful to Francesca Borzumati and Paolo Gambino 
for providing me with parts of their FORTRAN codes. 

\appendix
\section{Uncertainties from perturbative QCD}
Writing the perturbative expansion of some quantity in its general form for
an arbitrary gauge group, it can easily be decomposed into separately gauge 
invariant parts. Table~\ref{tab:coeff} shows for some (related) examples that 
after removing the group theoretical prefactors, all coefficients, $y_i$, are 
{\em strictly\/} of order unity, and that their mean, $\bar{y}$, is very close 
to zero. In particular, there is no sign of factorial growth of coefficients.
These observations offer a valuable tool to estimate the uncertainties
associated with the truncation of the loop expansion, so I would like 
to make them more precise. 

Assume (for simplicity) that the $y_i$ are random draws from some normal 
distribution with unknown mean, $\mu$, and variance, $\sigma^2$. One can show 
that the marginal distribution of $\mu$ follows a Student-t distribution with
$n-1$ degrees of freedom, $t_{n-1}$, centered about $\bar{y}$, and with 
standard deviation,
\be
  \Delta\mu = \sqrt{\sum_i (y_i - \bar{y})^2 \over n (n - 3)}.
\ee
As can be seen from the Table, $\mu$ is consistent with zero in all cases,
justifying the nullification of the unknown coefficients from higher loops. 
I next assert that the distribution of $\sigma$, conditional on $\mu = 0$, 
follows a scaled inverse-$\chi^2$ distribution with $n$ degrees of freedom, 
from which I obtain the estimate,
\be
   \sigma = \sigma_0 \pm \Delta\sigma 
   = \sqrt{\sum_i y_i^2\over n-2} \left[ 1 \pm \sqrt{1\over 2(n-4)} \right].
\ee
Inspection of the Table shows indeed that $\sigma$, as the typical size of 
a coefficient, is estimated to be $\lsim {\cal O}(1)$.

I now focus on the partial hadronic $Z$ decay width. As discussed in
Section~\ref{sec:zdecays}, the ${\cal O} (\alpha_s^3)$ term, $d_2$, is much
smaller than the $\pi^2$ term arising from analytical continuation. This is 
specifically true for the relevant case of $n_f = 5$ active flavors, where 
large cancellations occur between gluonic and fermionic loops. Notice, that 
the $D$-function, in contrast to $R_{\rm had}$, has opposite signs in 
the leading terms proportional to $C_A^2 C_F$ and $C_A C_F T_F n_F$. Indeed, 
the Adler $D$-function and the $\beta$-function have similar structures 
regarding the signs and sizes of the various terms (see Table~\ref{tab:coeff}),
and we {\em do expect\/} large cancellations in the $\beta$-function. 
The reason is that it has to vanish identically in the case of $N=4$ 
supersymmetry. Ignoring scalar contributions this case can be mimicked by 
setting $T_F n_f = 2 C_A$ (there are 2 Dirac fermions in the $N=4$ gauge 
multiplet) or $n_f = 12$ for QCD, which is of the right order. In fact, all 
known QCD $\beta$-function coefficients become very small for some value of 
$n_f$ between 6 and 16. We therefore have a reason to expect that similar 
cancellations will reoccur in the $d_i$ at higher orders. As a $1\sigma$ error 
estimate for $d_3$, I suggest to use the largest known coefficient 
($3\times 0.71$) times the largest group theoretical prefactor in the next 
order ($C_A^3 C_F$) which results in 
\be
\label{eq:d3estimate}
  d_3 = 0 \pm 77.
\ee
With Eq.~(\ref{eq:qcdns}) and $\hat\alpha_s(M_Z) = 0.120$ one can absorb all 
higher order effects into the ${\cal O}(\alpha_s^4)$-coefficient of 
$R_{\rm had}$, $r_3^{\rm eff} = - 81 \pm 63$. This shifts the extracted 
$\alpha_s$ from the $Z$ lineshape by +0.0005 and introduces the small 
uncertainty of $\pm 0.0004$.

The argument given above does certainly not apply to the {\em quenched\/} case,
$n_f = 0$, and indeed $d_2(n_f=0)$ is about $-73\%$ of the $\pi^2$ term, i.e., 
large and positive. In the case of $n_f = 3$, which is of interest for 
the precision determination of $\alpha_s$ from $\tau$ decays, $d_2$ is about 
$-38\%$ of the $\pi^2$ term. If one assumes that the same is true of $d_3$, one
would obtain $d_3(n_f=3) = 60$. Estimates based on the principles of minimal 
sensitivity, PMS, or fastest apparent convergence, FAC, yield 
$d_3(n_f=3) = 27.5$~\cite{Kataev95} so there might be some indications for
a positive $d_3(n_f=3)$. In any case, all these estimates lie within 
the uncertainty in Eq.~(\ref{eq:d3estimate}) and we will have to await 
the proper calculation of the ${\cal O}(\alpha_s^4)$-coefficient to test these 
hypotheses. Note, that the current $\tau$ decay analysis by the ALEPH 
Collaboration uses $d_3 = 50 \pm 50$~\cite{Barate98} which is more optimistic.

The analogous error estimate for the five-loop $\beta$-function coefficient
yields,
\be
\label{eq:beta4}
  \beta_4 = 0 \pm 579.
\ee
To get an estimate for the uncertainty in the RG running of $\hat\alpha_s$,
I translate Eq.~(\ref{eq:beta4}) into
\be
  \beta_3 = \beta_3 \pm {\hat\alpha_s(\mu_0)\over\pi} \beta_4,
\ee
where $\mu_0$ is taken to be the lowest scale involved. This overestimates
the uncertainty from $\beta_4$, thereby compensating for other neglected terms 
of ${\cal O}(\alpha_s^{n+4} \ln^n \mu^2/\mu_0^2)$. For the RG evolution from 
$\mu = m_\tau$ to $\mu = M_Z$ this yields an uncertainty of 
$\Delta\alpha_s(M_Z) = \pm 0.0005$. 
Conversely, for fixed $\alpha_s(M_Z) = 0.120$, I obtain 
$\hat\alpha_s(\hat{m}_b) = 0.2313 \pm 0.0006$, 
$\hat\alpha_s(m_\tau) = 0.3355\pm 0.0045$, and 
$\hat\alpha_s(\hat{m}_c) = 0.403 \pm 0.011$, where I have used 
$\hat{m}_b = 4.24$~GeV and $\hat{m}_c = 1.31$~GeV. For comparison, the ALEPH 
Collaboration quotes an evolution error of $\Delta\alpha_s(M_Z) = \pm 0.0010$
which is twice as large. I emphasize that it is important to adhere to 
consistent standards when errors are estimated. This is especially true in 
the context of a global analysis where the precisions of the observables enter 
as their relative weights.

\end{document}